\documentclass[preprint,12pt]{elsarticle}
\usepackage[breaklinks=true]{hyperref}
\usepackage{soul,xcolor,bm}
\usepackage{geometry}
\usepackage{multirow}
\usepackage{amsmath}
\biboptions{numbers,sort&compress}

\usepackage{amssymb}

\journal{arXiv}

\begin{document}

\begin{frontmatter}



\title{Finite-Temperature Atomistic and Continuum Stress Fields of Coherent Precipitates with a Small Lattice Misfit}

\author[inst1]{Anas Abu-Odeh}
\author[inst1]{James Warren}

\affiliation[inst1]{organization={National Institute of Standards and Technology},
            city={Gaithersburg},
            postcode={20899}, 
            state={MD},
            country={USA}}

\begin{abstract}
An accurate description of elastic effects of coherent microstructures is necessary for the predictive modeling of microstructural evolution in many structural materials. To date, there has not been a demonstration on how continuum elasticity models are able to reproduce finite-temperature stress-fields and elastic energy estimates of coherent precipitates from atomistic simulations. We present a comparison of stress-fields of coherent precipitates in the body-centered cubic (BCC) Fe-Cr system obtained from atomistic simulation data and from continuum elasticity modeling. The magnitude and topology of the stress-fields show good agreement between the two approaches, and we show the importance of elastic effects on the Gibbs-Thompson effect for this small lattice misfit system. We conclude with a discussion of potential complications of continuum modeling for systems with larger misfit.
\end{abstract}

\end{frontmatter}


\section{Introduction}

The microstructure of a mixture of coherent phases is heavily influenced by elastic effects \cite{Voorhees2004}. These can be due to compositionally-dependent lattice parameters which can cause a precipitate of one composition to be strained in a matrix of another composition. As these elastic effects influence processes such as nucleation \cite{Clouet2009} and precipitate shape evolution \cite{Thompson1994}, an accurate description of the elastic fields and elastic energy is necessary for the predictive modeling of microstructure evolution.

While the importance of elastic effects on microstructure is well understood and has been a motivation for many previous studies, most of these studies have focused on using or developing continuum models that are usually compared to other continuum models or numerical techniques. For example, Moulinec and Suquet \cite{Moulinec1998} developed an iterative spectral approach to determine the elastic fields of a given microstructure and compared their results to analytical solutions. Hu and Chen \cite{Hu2001} also developed an iterative spectral approach to determine elastic displacement fields in a continuous medium with diffuse interfaces and compared the results to a conjugate gradient approach. Simon et al. \cite{Simon2020} has made a comparison of diffuse and sharp interface calculations of elastic fields of misfitting inclusions. While these studies are important, the comparisons are done between models or techniques that share similar levels of approximation. In general, comparisons of the elastic field of an inclusion/inhomogenienty against higher fidelity modeling, such as classical atomistic modeling, are rare.

One exception is a study by Hoang et al. \cite{Hoang2011} where atomistic stress fields and displacements were evaluated at 0 K for both inclusions and inhomogeneities and compared to elasticity theory using isotropic elastic constants. Another exception is a study by Boussinot et al. \cite{Boussinot2010} where a phase-field model coupled with elasticity was found to qualitatively reproduce the shape evolution of a two-dimensional inhomogeneity with either isotropic or anisotropic elastic constants when compared to previous on-lattice Monte Carlo simulations. While these studies show that continuum elasticity theory can reproduce results from some lower length scale modeling, an important question that remains is whether or not elasticity theory can reproduce results from finite-temperature atomistic simulations, especially when compositional fluctuations are present.

In this paper, we aim to show how continuum modeling can be used to reproduce finite-temperature stress fields obtained from atomistic simulations of a coherent precipitate with a small lattice misfit with respect to the matrix. We accomplish this by obtaining finite-temperature inputs from atomistic data that feed into the continuum modeling. In Section \ref{methods}, we discuss the methods to obtain atomistic data and how the continuum model is evaluated. Section \ref{results} presents results for various inputs into the continuum model, comparisons of stress fields between atomistic simulations and the continuum model, estimates of total elastic energies of the systems, as well as an example of the importance of elastic effects when estimating interfacial free energies. Section \ref{discussion} discusses future challenges for systems with larger lattice misfit, and we conclude with Section \ref{conclusion}.

\section{Methods} \label{methods}

All simulations, atomistic and continuum, are carried out with fully periodic boundary conditions.

\subsection{Fe-Cr Interatomic Potential}

We choose to work with the body-centered cubic (BCC) Fe-Cr system due to the relatively small misfit between Fe and Cr. To represent this system, we modify an existing concentration-dependent EAM (CD-EAM) potential for the BCC Fe-Cr system \cite{Stukowski2009}. This is done by modifying the concentration dependent function ($h(x)$ in \cite{Stukowski2009}) in order to reduce the peak in the heat of mixing curve for random Fe-Cr solid solutions. We do this as the evaluated phase diagram for this potential \cite{Sadigh2012_int} has an unrealistically low solubility of Fe in Cr at high temperatures. There exist multiple Fe-Cr interatomic potentials, with the most promising one being the two-band model (TBM) extension of the embedded-atom method (EAM) by Eich et al. \cite{Eich2015}. While this potential is able to closely reproduce the metastable BCC miscibility gap when compared to experimentally assessed phase diagrams, the current implementation of the Monte Carlo procedure that we use in the Large-scale Atomic/Molecular Massively Parallel Simulator (LAMMPS) software \cite{Plimpton1995, Sadigh2012} is computationally inefficient when using a TBM-EAM potential. Our modified CD-EAM potential allows for higher solubility of Fe in Cr, but still exhibits phase-separation above the critical temperature of the TBM-EAM potential. This may be due to a lack of flexibility in the CD-EAM formalism compared to the TBM-EAM formalism. Regardless, we use our modified CD-EAM potential as a compromise between an accurate description of the Fe-Cr system and a reduced computational cost of Monte Carlo moves in LAMMPS. The parameters of the new CD-EAM potential and its heat of mixing curve are given in page 1 of the Supplementary Material.

\subsection{Hybrid Monte Carlo/Molecular Dynamics} \label{hybridmcmd}

Hybrid Monte Carlo/molecular dynamics (MC/MD) simulations are carried out in the isobaric semi-grand canonical (SGC) or the isobaric variance-constrained semi-grand canonical (VC-SGC) ensembles. One MC sweep (which we define as equal to having a number of MC attempts equal to the number of total atoms in the system, $N$) is carried out after every ten MD time steps, where one time step is 2.5 fs. MD is carried out in the $NPT$ ensemble, with an isotropic deformation of the simulation box coupled to a barostat targeting zero pressure. In order to take advantage of an efficient implementation of Monte Carlo for the CD-EAM potential, we use the $\mathbf{vcsgc-lammps}$ package from Ref. \cite{vcsgc} with the 3 Mar 2020 version of LAMMPS. For a binary mixture of A and B atoms, SGC MC relates the compositional dependence of the difference in chemical potentials of B and A ($\Delta\mu(c) = \mu_B (c) - \mu_A (c)$) to the derivative of the Gibbs free energy per atom with respect to composition, $\partial g / \partial c$, through: 
\begin{equation} \label{eq:1}
    \frac{\partial g}{\partial c} \bigg|_{c} = \Delta\mu(c)
\end{equation}
where $c$ is the concentration of B atoms. When SGC MC is used for hybrid MC/MD, we select various values of $\Delta\mu$, run simulations for 6,000 MC sweeps, measure the average composition $\langle c\rangle$ of the last 4,000 MC sweeps, and then map out $\Delta\mu$ as a function of $c$.

We carry out MC in the VC-SGC ensemble for two reasons: to check that the different ensembles give identical results in the single phase regions and to efficiently sample a two phase mixture. There are three input parameters when running MC in the VC-SGC ensemble using spatial domain decomposition in LAMMPS: the average constraint parameter, $\Delta\mu^v$, the variance constraint parameter, $\kappa$, and the target concentration, $c_t$. When sampling equilibrium states in this ensemble, these parameters and the measured average concentration of the states are related to $\partial g / \partial c$ through \cite{Sadigh2012}:
\begin{equation} \label{eq:2}
    \frac{\partial g}{\partial c} \bigg|_{c} = \Delta\mu^v +2\kappa N (\langle c \rangle - c_t).
\end{equation}
The choice of $\Delta\mu^v$ can dramatically impact the acceptance rate of MC moves, with the optimal choice being $\Delta\mu^v = \Delta\mu(c_t)$ \cite{Sadigh2012}. As $\Delta\mu(c_t)$ is not known beforehand, we use a similar feedback approach to that of Ref. \cite{Mishin2014}. For a number of MC sweeps, the value of $\Delta\mu^v$ is updated after each MC sweep according to:
\begin{equation} \label{eq:3}
    \Delta\mu^v_i = \Delta\mu^v_{i-1} - \eta \Bigl(\frac{c_{i-1}+c_{i-2}}{2} - c_t \Bigr)
\end{equation}
where $i$ is the current MC sweep and $\eta$ is a positive constant. An initial value of $\Delta\mu^v$ was found using a crude estimate of the value of $\Delta\mu$ at coexistence from cheap SGC MC/MD simulations of small systems. $\kappa$ is set to 10 for all MC sweeps. We note that LAMMPS rescales the set value of $\kappa$ by multiplying it by $k_B T/N$, where $k_B$ is the Boltzmann constant, and $T$ is the temperature. When running VC-SGC MC/MD simulations along with the feedback approach, for the first 500 MC sweeps $\eta$ was set to 0.1 and for the next 500 MC sweeps $\eta$ was set to 0.01. Afterwards the feedback approach is not used ($\eta$ = 0) and the value of $\Delta\mu^v$ remains  constant. VC-SGC MC/MD simulations are carried out for 5000 MC sweeps with this value of $\Delta\mu^v$, and the average concentration is obtained from the last 4000 MC sweeps.

We note that while Eq. \ref{eq:1} is a fundamental relationship from thermodynamics, and that Eqs. \ref{eq:2} and \ref{eq:3} are taken from Refs. \cite{Sadigh2012,Mishin2014}, \textit{these equations must be modified} when using LAMMPS due to details of the implementation of the SGC and VC-SGC ensembles. We discuss these modifications in pages 2 and 3 of the Supplementary Material.

\subsection{Modeling the Free Energy of BCC Fe-Cr}

The information from the SGC and VC-SGC MC/MD simulations will aid in determining the coexistence compositions, which are needed in order to obtain finite-temperature elastic constants for each phase. Determining the coexistence compositions requires estimating the Gibbs free energy as a function of composition. We model the Gibbs free energy per atom of the BCC phases through:
\begin{equation} \label{eq:4}
    g(c) = g_{Fe} + c(g_{Cr}-g_{Fe}) + k_B T [(c)\mathrm{ln}(c) + (1-c)\mathrm{ln}(1-c)] + g^{ex}(c)
\end{equation}
where $g_{Fe}$ and $g_{Cr}$ are the Gibbs free energy per atom of pure Fe and Cr, and $g^{ex}(c)$ represents the excess Gibbs free energy contribution (which is equal to zero in the pure states). The term with square brackets represents the ideal entropy contribution. We model the excess term using a Redlich-Kister polynomial \cite{Redlich1948}:
\begin{equation} \label{eq:5}
    g^{ex}(c) = c(1-c) \sum_{i = 0}^n L_i (1-2c)^i
\end{equation}
where $n+1$ is the number of terms in the polynomial, and $L_i$ are coefficients fit to MC results. From Eq, \ref{eq:4}, $\partial g/ \partial c$ is given as:
\begin{equation} \label{eq:6}
    \frac{\partial g}{\partial c} \bigg|_{c} = g_{Cr} - g_{Fe} + k_B T \mathrm{ln}\Bigl(\frac{c}{1-c} \Bigr) + \frac{\partial g^{ex}}{\partial c} \bigg|_{c}
\end{equation}
where the left-hand side is obtained from MC/MD simulations. The $\frac{\partial g^{ex}}{\partial c}$ term can be fit to the derivative of Eq. \ref{eq:5} with respect to composition, but in order to do so $g_{Cr}$ and $g_{Fe}$ must be determined. These values are determined using the nonequilibrium thermodynamic integration method of Ref. \cite{Freitas2016}. We summarize the procedure here. For a given temperature, the equilibrium lattice constant of the material is obtained through $NPT$ MD, the mean-squared displacement from $NVT$ MD at the equilibrium lattice constant is used to obtain an effective spring constant, and the effective spring constant is used to parameterize an Einstein crystal reference state. The work required to switch to and from this reference state allows for an accurate estimate of the Helmholtz free energy of a pure material, as described in Ref. \cite{Freitas2016}. When the volume is fixed such that pressure is zero, this is equivalent to the Gibbs free energy at zero pressure. We use a supercell with 40x40x40 replicated BCC conventional unit cells (128,000 atoms) for these calculations, and during the switching procedure we use an equilibration period of 50,000 time steps, and a switching period of 200,000 time steps, where each time step is 1 fs. We repeat the switching procedure using five independent samples with different randomly initialized velocities to determine an average free energy and an estimate of error.

\subsection{Finite-Temperature Elastic Constants}

Parameterizing the continuum elasticity model includes obtaining finite-temperature elastic constants. To do this, we use the stress-fluctuation method as implemented in LAMMPS \cite{Clavier2023}. For these calculations, we run MD in the $NVT$ ensemble for 200,000 time steps of 1 fs each using the 2 Aug 2023 version of LAMMPS. At each time step, the second derivative of the potential energy per unit volume with respect to strain ($C_{ijkl}^B$) is measured along with the instantaneous virial stress ($\sigma_{ij}^B$). The finite-temperature elastic constants ($C_{ijkl}$) are given by:
\begin{equation} \label{eq:7}
    C_{ijkl} = \langle C_{ijkl}^B \rangle - \frac{V}{k_B T} [\langle \sigma_{ij}^B \sigma_{kl}^B \rangle - \langle \sigma_{ij}^B \rangle \langle \sigma_{kl}^B \rangle] + \rho k_B T (\delta_{ik} \delta_{jl} + \delta_{il} \delta_{jk})
\end{equation}
where $V$ is the volume of the system, $\langle \rangle$ denote averaged quantities, $\rho$ is the atomic density, and $\delta_{ik}$ is the Kronecker delta. Five independent configurations with 128,000 atoms each at coexistence compositions for each temperature are generated from hybrid VC-SGC MC/MD simulations. The values of $C_{ijkl}$ are evaluated for each configuration and then averaged together.

\subsection{Extracting Composition and Stress Fields from Atomistic Data} \label{averaging}

\begin{figure}[h!]
    \centering
    \includegraphics[width=1\textwidth]{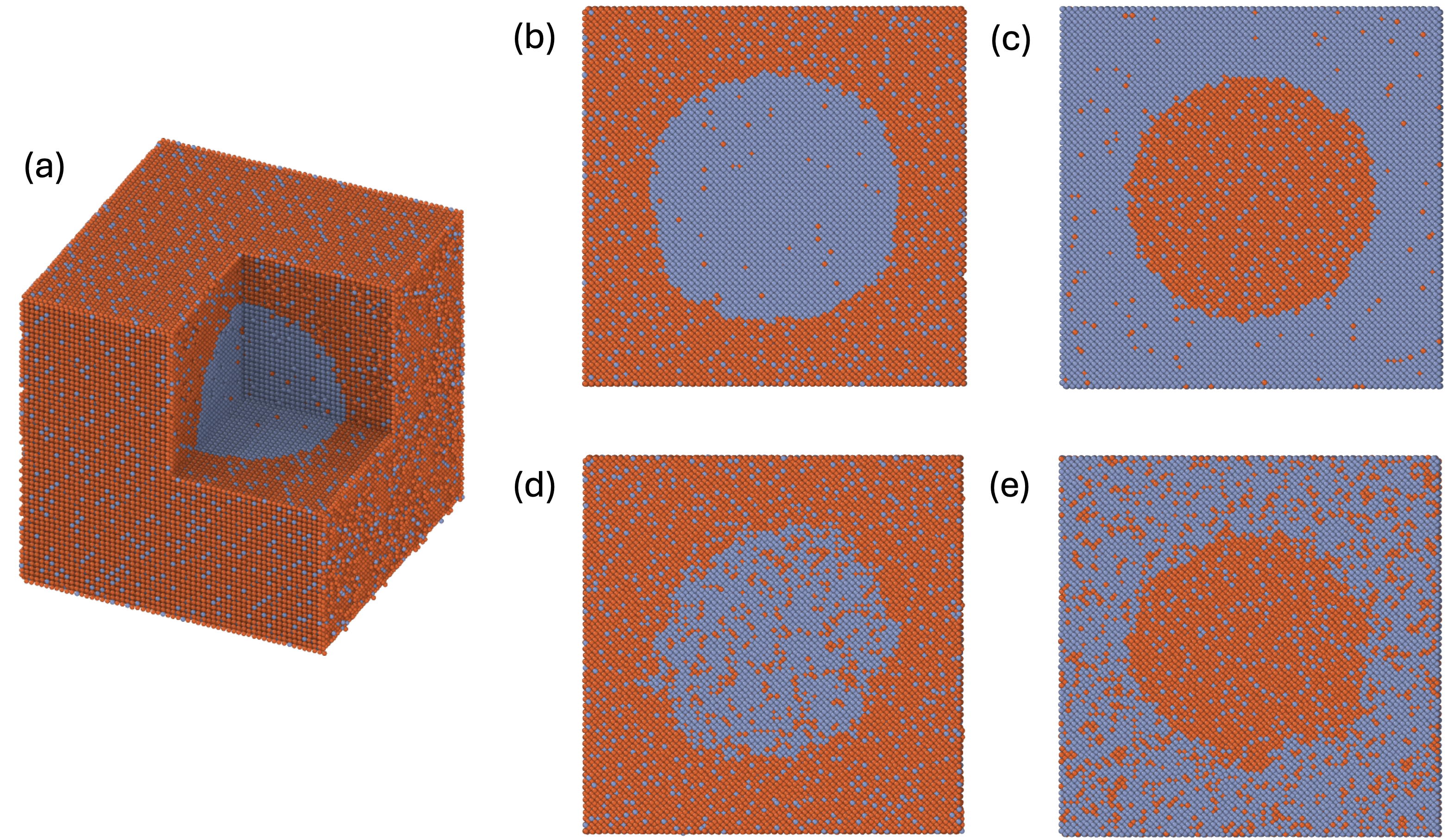}
    \caption{Atomistic snapshots from VC-SGC MC/MD simulations. (a) An example of a snapshot of a supercell of an $\alpha'$ precipitate in an $\alpha$ matrix at 600 K with a corner removed to show the precipitate. Shown in (b-e) are atoms taken from a two-dimensional slice of a spherical precipitate in a matrix. (b) Shows a Cr-rich phase in an Fe-rich matrix at 600 K and (c) shows a Fe-rich phase in a Cr-rich matrix at 600 K. (d) and (e) are similar to (b) and (c), except at 1000 K.}
    \label{fig:snapshots}
\end{figure}

Two-phase mixtures composed of a spherical precipitate in a matrix are generated through hybrid VC-SGC MC/MD using supercells of 60x60x60 replicated BCC conventional unit cells (432,000 atoms). At any given snapshot (see for example Figure \ref{fig:snapshots}), the composition and stress fields will have significant noise due to local compositional inhomogeneities, as well as variance due to instantaneous displacements and velocities. To facilitate comparison with the continuum model, a smoother, averaged set of fields is required. A snapshot from a previous VC-SGC MC/MD simulation is used to initialize another simulation of 50,000 time steps (5,000 MC sweeps). After every 100 time steps (or 10 MC sweeps), a snapshot is generated with time averaged atomic positions, stresses, and concentrations over the last 100 time steps. Each of the 500 snapshots are centered by bringing the center of mass of the primary species of the precipitate to the center of the simulation box while accounting for periodic boundary conditions \cite{Bai2008}. The data for each snapshot is binned in a 60x60x60 array, and the arrays are averaged over every snapshot. As the atomic stresses output by LAMMPS are in units of stress$\times$volume, the stress field is divided by the effective unit cell volume, which we define as the average unit cell volume of the full system.

\subsection{Estimate of Interfacial Width}

To have a better representative compositional field for the continuum model, an interfacial width needs to be estimated. A supercell of 60x60x60 replicated BCC conventional unit cells (432,000 atoms) is created with a slab of Fe atoms on one side and a slab of Cr atoms on the other. Hybrid VC-SGC MC/MD is carried out according to the procedure in section \ref{hybridmcmd} to equilibrate a structure with approximately equal phase fractions of each phase. The compositional profile was binned in the direction normal to the interface and fit to a hyperbolic tangent profile:
\begin{equation} \label{eq:8}
    c(x) = \frac{c_1+c_2}{2} + (c_1-\frac{c_1+c_2}{2}) \mathrm{tanh}\Bigl(\frac{x-x_c}{l}\Bigr)
\end{equation}
where $c_1$ and $c_2$ are the compositions of the two phases, $x_c$ is the location of the interface, and $l$ is a fitting parameter. The interfacial profiles and values of $l$ are presented on page 4 of the Supplementary Material.

\subsection{Continuum Elasticity Model}

We use an elasticity model for inhomogeneous microstructures that is commonly used in phase-field models \cite{Hu2001,Boussinot2010}. In the case of zero applied stress, the elastic energy ($E^{el}$) of a system in the framework of linear elasticity is given by:
\begin{equation} \label{eq:9}
    E^{el} = \frac{1}{2} \int d\vec{r} C_{ijkl}(\vec{r})\{\epsilon_{ij}(\vec{r}) - \epsilon^0_{ij}(\vec{r})\}\times\{\epsilon_{kl}(\vec{r}) - \epsilon^0_{kl}(\vec{r})\}
\end{equation}
where $\epsilon_{ij}$ is the total strain, $\epsilon^0_{ij}$ is the stress-free eigenstrain, and the difference of the two corresponds to the elastic strain. The elastic constants are dependent on the local composition field ($c(\vec{r})$), which varies in space. We assume that the elastic constants vary with composition as:
\begin{equation} \label{eq:10}
    C_{ijkl} = \bar{C}_{ijkl} + C'_{ijkl}\Delta c(\vec{r})
\end{equation}
where $\bar{C}_{ijkl}$ is the average elastic constant tensor given by:
\begin{equation} \label{eq:11}
    \bar{C}_{ijkl} = \frac{C^{\alpha}_{ijkl}(c^{\alpha'}_{eq} - \bar{c}) + C^{\alpha'}_{ijkl}(\bar{c}-c^\alpha_{eq})}{c^{\alpha'}_{eq} - c^{\alpha}_{eq}}
\end{equation}
where $\bar{c}$ is the average composition of the entire system, and $C^{\alpha}_{ijkl}$ and $C^{\alpha'}_{ijkl}$ are the elastic constants for the Fe-rich and Cr-rich phase, respectively, at their coexistence compositions ($c^{\alpha}_{eq}$ and $c^{\alpha'}_{eq} $). $\Delta c(\vec{r})$ is the difference between the local concentration and the average concentration. $C'_{ijkl}$ represents the inhomogenous contribution to the elastic constants and is given by:
\begin{equation} \label{eq:12}
    \bar{C}'_{ijkl} = \frac{C^{\alpha'}_{ijkl} - C^{\alpha}_{ijkl}}{c^{\alpha'}_{eq} - c^{\alpha}_{eq}}.
\end{equation}
The compositionally dependent eigenstrain, $\epsilon^0_{ij}(\vec{r})$, in the purely dilational case is approximated by:
\begin{equation} \label{eq:13}
    \epsilon^0_{ij}(\vec{r}) = \frac{a^{\alpha'}-a^{\alpha}}{\bar{a}(c^{\alpha'}_{eq} - c^{\alpha}_{eq})}\delta_{ij}\Delta c(\vec{r})
\end{equation}
where $a^{\alpha}$ and $a^{\alpha'}$ are the lattice constants of the Fe-rich and Cr-rich phases, respectively, at the coexistence compositions, and $\bar{a}$ is the effective lattice constant of the supercell with the two-phase mixture. In this work, the lattice constants, coexistence compositions, and elastic constants are all obtained from atomistic simulations as described above.

A convenient way to represent the total strain, $\epsilon_{ij}(\vec{r})$, is to decompose it as:
\begin{equation} \label{eq:14}
    \epsilon_{ij}(\vec{r}) = \bar{\epsilon}_{ij} + \Delta \epsilon_{ij}(\vec{r})
\end{equation}
where $\bar{\epsilon}_{ij}$ is a homogeneous average strain and $\Delta \epsilon_{ij}(\vec{r})$ is a local heterogeneous strain. The heterogeneous strain is related to the local displacements, $u(\vec{r})$, through:
\begin{equation} \label{eq:15}
    \Delta\epsilon_{ij}(\vec{r}) = \frac{1}{2}\Bigl[\frac{\partial u_i(\vec{r})}{\partial r_j} + \frac{\partial u_j(\vec{r})}{\partial r_i}  \Bigr].
\end{equation}
The resulting stress field, $\sigma_{ij}(\vec{r})$, is given by:
\begin{equation} \label{eq:16}
    \sigma_{ij}(\vec{r}) = C_{ijkl}(\vec{r})[\epsilon_{kl}(\vec{r})-\epsilon^0_{kl}(\vec{r})].
\end{equation}
To determine the stress field and the elastic energy of a system at mechanical equilibrium, the following equations must first be solved to determine the values of $u_i(\vec{r})$ and $\bar{\epsilon}_{ij}$ that minimize Eq. \ref{eq:9}:
\begin{equation} \label{eq:17}
    \frac{\delta E^{el}}{\delta u_i (\vec{r})} = 0
\end{equation}
\begin{equation} \label{eq:18}
    \frac{\delta E^{el}}{\delta \bar{\epsilon}_{ij}} = 0
\end{equation}
where the last equation is equal to zero as we have zero applied stress in our systems. Eqs. \ref{eq:17} and \ref{eq:18} are solved using an iterative spectral method as described in Refs. \cite{Hu2001,Boussinot2010}. We found that we needed five iterations to obtained converged results for our systems, and that using a regular grid spacing of half the lattice constant (a grid of 120x120x120 points) resulted in a well converged elastic energy when compared to finer grid sizes.

As a composition field is required to evaluate the above equations, we estimate the composition field as a spherical precipitate and a surrounding matrix that are at their respective coexistence compositions, with an interfacial profile given by Eq. \ref{eq:8}. The radius of the precipitate is chosen so that the global composition of the continuum field is near that of the global composition of the atomistic fields used for comparison. The chosen radii are reported in the Results section.

\section{Results} \label{results}

\subsection{Determining Coexistence Compositions}

\begin{figure}[h!]
    \centering
    \includegraphics[width=1\textwidth]{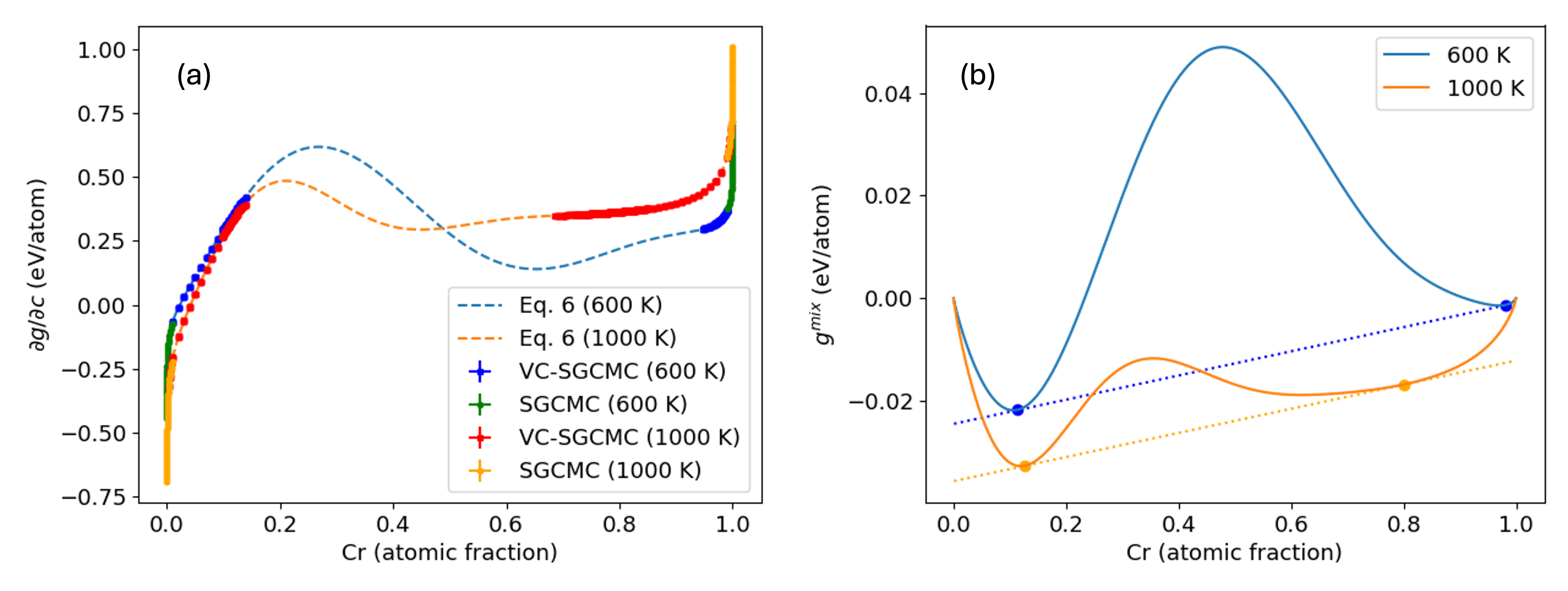}
    \caption{(a) Data points are results from MC/MD simulations in either the SGC MC or VC-SGC MC ensemble relating $\partial g/ \partial c$ to the average composition and dashed lines are obtained from Eq. \ref{eq:6} where the excess term has been fit using a Redlich-Kister polynomial. (b) Solid lines are mixing free energies obtained from the ideal and excess terms in Eq. \ref{eq:4}, dotted lines are obtained from the common tangent construction, and the data points represent the coexistence compositions.}
    \label{fig:fecr_mc}
\end{figure}

Coexistence compositions are obtained through a common tangent construction using the free energy function given by Eq. \ref{eq:4}. To use Eq. \ref{eq:4}, the free energies of the pure states are needed. The values of $g_{Fe}$ and $g_{Cr}$ at the temperatures of 600 K and 1000 K are given in Table \ref{tab:end_member}. Additionally, an estimate for the excess term, $g^{ex}(c)$, must be made. This can be done by fitting Eq. \ref{eq:6} to MC/MD results of $\partial g/ \partial c$. Figure \ref{fig:fecr_mc} (a) shows the MC/MD results as well as values using Eq. \ref{eq:6} where the derivative of a Redlich-Kister polynomial is used to describe $\frac{\partial g^{ex}}{\partial c}$. For 600 K, the polynomial included 7 terms, while for 1000 K, the polynomial included 10 terms. The values of $L_i$ are given in the Supplementary Material. The VC-SGC MC results are obtained using a range of supercells from 20$\times$20$\times$20 to 60$\times$60$\times$60 replicated BCC conventional unit cells (16,000 to 432,000 atoms) as smaller supercells allow for sampling more supersaturated states of the phases. The SGC MC results are obtained using a supercell with 40$\times$40$\times$40 replicated BCC conventional unit cells (128,000 atoms). 

\begin{table}[h!]
    \centering
	\begin{tabular}{l | l | l }
        \multirow{2}{*}{} 
         & 600 K & 1000 K \\ \hline && \\
		$g_{Fe}$ (eV/atom) & -4.242902(4) & -4.461397(5)  \\ && \\ \hline && \\
		$g_{Cr}$ (eV/atom) & -3.926012(3) & -4.121260(6)  \\ && \\

    \end{tabular}
	\caption{\label{tab:end_member} Average free energies of pure Fe and Cr obtained from a nonequilibrium thermodynamic integration method of five independent runs at the temperatures of 600 K and 1000 K. Values in parentheses are estimates of two times the standard error.}
\end{table}

Once we have estimated the coefficients of the Redlich-Kister polynomial that describe $g^{ex}(c)$, we can use Eq. \ref{eq:4} to construct a common tangent and obtain the coexistence compositions. This is shown in Gibbs free energy of mixing plot in Figure \ref{fig:fecr_mc} (b). From this construction we take the coexistence compositions at 600 K to be at Cr atomic fractions of 0.1134 and 0.982, and at 1000 K to be at Cr atomic fractions of 0.1272 and 0.801. 

\subsection{Lattice and Elastic Constants}

\begin{figure}[h!]
    \centering
    \includegraphics[width=0.75\textwidth]{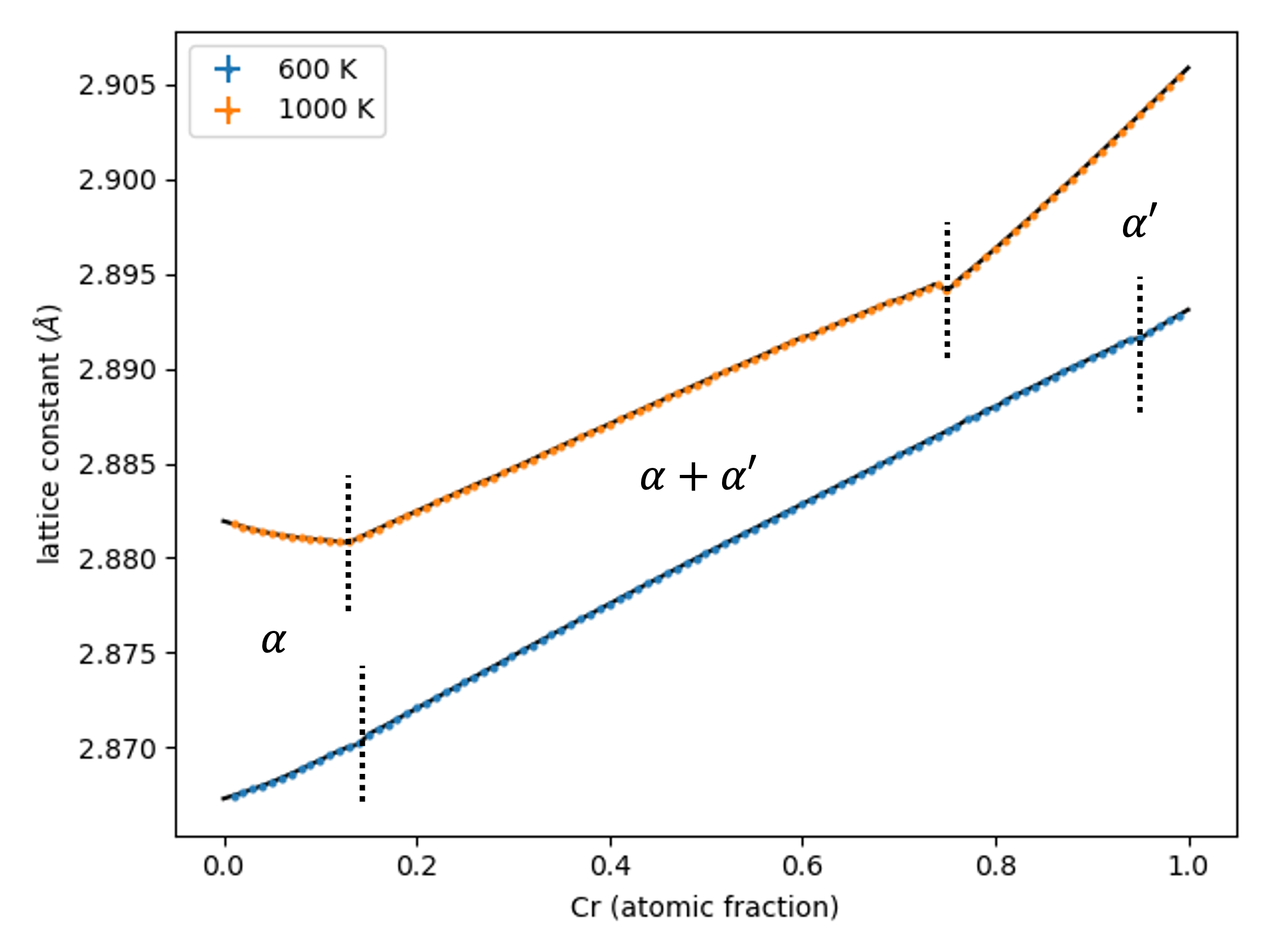}
    \caption{Average lattice constant obtained from hybrid VC-SGC MC/MD results at a system size of 432,000 atoms. The blue data points are from simulations at 600 K, and the orange data points are from simulations at 1000 K. The solid black lines are spline interpolation curves, and the dotted black lines demarcate regions of single phases from the region of a two-phase mixture. The dotted black lines do not coincide with the compositions at coexistence.}
    \label{fig:lattice_constant}
\end{figure}

The measured lattice constants from hybrid VC-SGC MC/MD simulations are given in Figure \ref{fig:lattice_constant}. In that figure, regions are separated based on whether the simulation box had a single phase or a two-phase mixture. From a spline fit to the simulation data, we find that the lattice constant of the $\alpha$ and $\alpha'$ phase at the coexistence compositions are 2.87 $\mathrm{\AA}$ and 2.893 $\mathrm{\AA}$, respectively, for 600 K, and are 2.881 $\mathrm{\AA}$ and 2.896 $\mathrm{\AA}$, respectively, for 1000 K. The effective lattice constant ($\bar{a}$) for the supercell of a two phase mixture was obtained using the spline fit.

The finite-temperature elastic constants at the coexistence compositions are reported in Table \ref{tab:elastic}. At both 600 K and 1000 K, the elastic constants of the $\alpha$ phase are lower than that of the $\alpha'$ phase.

\begin{table}[h!]
    \centering
	\begin{tabular}{l | l l | l l }
        \multirow{2}{*}{} 
         & 600 K && 1000 K \\ \hline && \\
         & $\alpha$ & $\alpha'$ & $\alpha$ & $\alpha'$ \\ \hline && \\
		$C_{11}$ (GPa) & 208.87(9) & 255.5(8) & 175.6(4) & 202.4(4)  \\ && \\ \hline && \\
		$C_{12}$ (GPa) & 109.5(1) & 119.7(2) & 101.5(3) & 111.4(3) \\ && \\ \hline && \\
            $C_{44}$ (GPa) & 83.01(7) & 118.66(6) & 76.82(5) & 100.0(2) \\ && \\ 

    \end{tabular}
	\caption{\label{tab:elastic} Finite-temperature elastic constants for the $\alpha$ and $\alpha'$ phases at their coexistence compositions for 600 K and 1000 K. The numbers in parentheses are two times the standard error.}
\end{table}

\subsection{Comparison of Stress Fields}

Instead of comparing the field of each stress component from atomistic data and continuum modeling, we choose to compare the pressure and von Mises stress fields as a measure of hydrostatic and deviatoric stresses. We evaluate the pressure ($P(\vec{r})$) field through:

\begin{equation} \label{eq:19}
    P(\vec{r}) = -\frac{\sigma_{11}(\vec{r})+\sigma_{22}(\vec{r})+\sigma_{33}(\vec{r})}{3}
\end{equation}
and the von Mises stress ($\sigma_{VM}(\vec{r})$) field through:
\begin{equation} \label{eq:20}
\begin{split}
    2\sigma_{VM}^2(\vec{r}) = (\sigma_{11}(\vec{r})-\sigma_{22}(\vec{r}))^2+(\sigma_{22}(\vec{r})-\sigma_{33}(\vec{r}))^2+(\sigma_{33}(\vec{r})-\sigma_{11}(\vec{r}))^2 
\\ +6(\sigma_{12}^2(\vec{r})+\sigma_{13}^2(\vec{r})+\sigma_{23}^2(\vec{r})).
\end{split}
\end{equation}
The atomistic fields were calculated from the average field of each component as described by Section \ref{averaging}. Once a pressure or von Mises field is calculated, we take advantage of the cubic symmetry of the BCC system and rotationally average the fields. We found that this helps to reduce the residual noise that still persists from the atomistic data.

\subsubsection{Pressure Fields}

\begin{figure}[h!]
    \centering
    \includegraphics[width=1\textwidth]{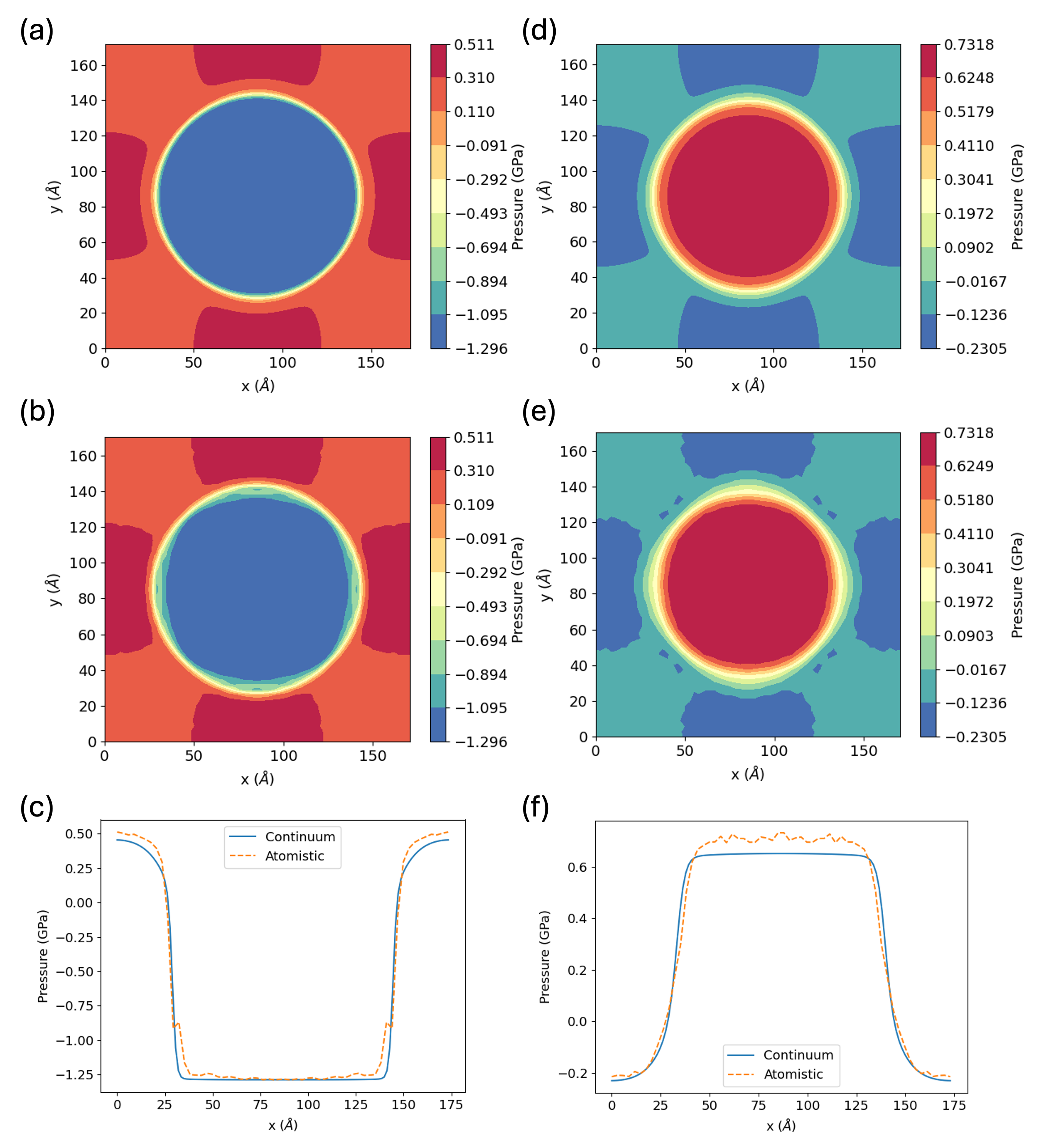}
    \caption{(a) Pressure field of an $\alpha$ precipitate in a $\alpha'$ matrix at 600 K from continuum modeling. (b) Same as (a) except from atomistic data. (c) Line scan across the center of the precipitate for (a) and (b). (d-f) Similar to (a-c), except for an $\alpha'$ precipitate in an $\alpha$ matrix at 1000 K.}
    \label{fig:press}
\end{figure}

Figure \ref{fig:press} shows a comparison of the pressure field between atomistic data and continuum modeling for two cases: an $\alpha$ precipitate in an $\alpha'$ matrix at 600 K (a-c), and an $\alpha'$ precipitate in an $\alpha$ matrix at 1000 K (d-f). The pressure field distributions are similar between the atomistic field and the continuum field. A similar magnitude of pressure is present in the precipitate, as well as near the edge of the simulation box. This latter effect is due to periodic boundary conditions causing the precipitates to interact with each other. Qualitative differences appear near the interface. These differences may be due to higher-order effects mentioned in the discussion section. A comparison of the pressure fields for an $\alpha$ precipitate in an $\alpha'$ matrix at 1000 K, and an $\alpha'$ precipitate in an $\alpha$ matrix at 600 K is given on page 6 of the Supplementary Material.

\subsubsection{von Mises Stress Fields}

\begin{figure}[h!]
    \centering
    \includegraphics[width=1\textwidth]{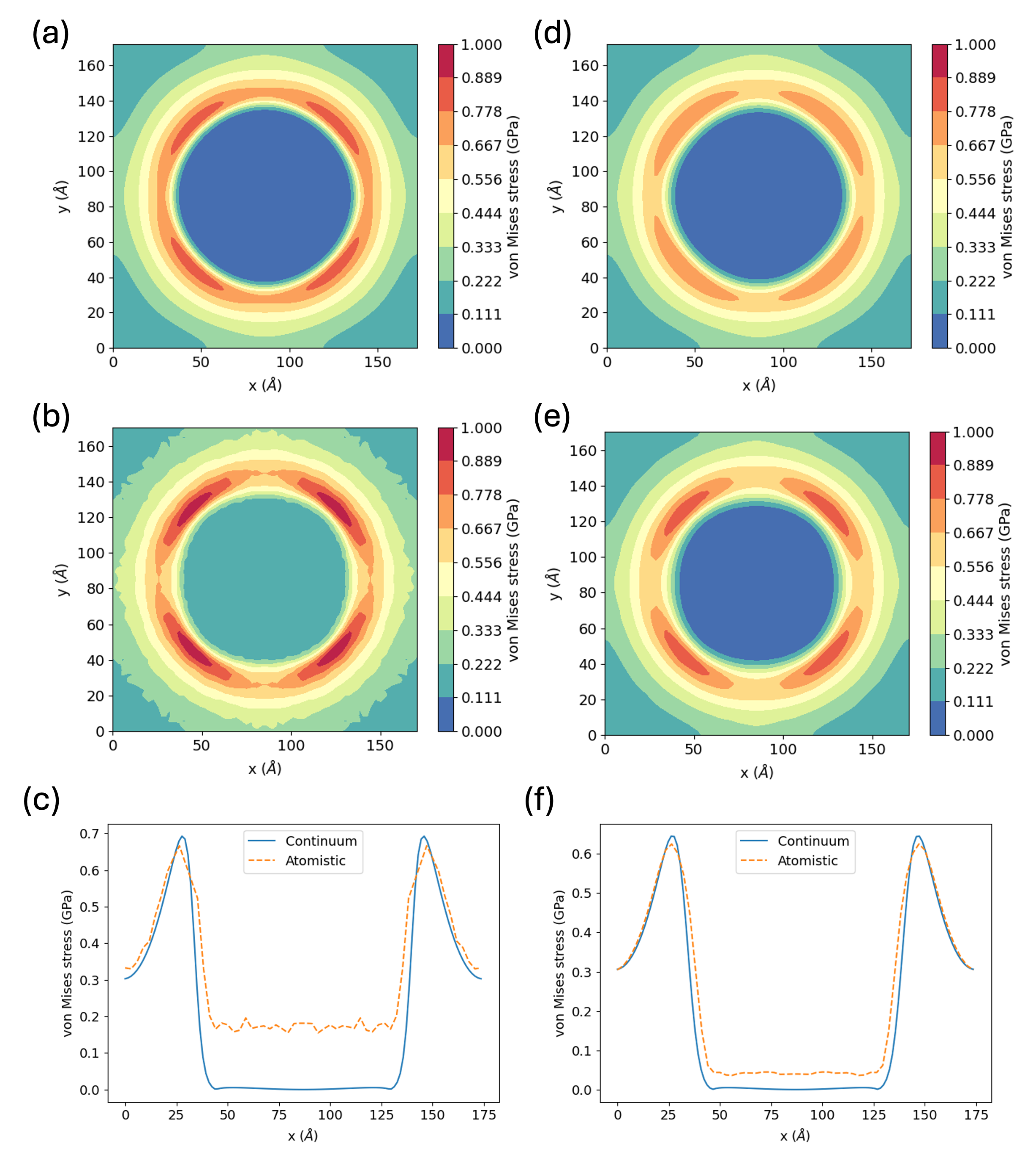}
    \caption{(a) von Mises stress field of an $\alpha'$ precipitate in an $\alpha$ matrix at 1000 K from continuum modeling. (b) Same as (a) except from atomistic data. (c) Line scan across the center of the precipitate for (a) and (b). (d-f) Similar to (a-c), except that a Gaussian filter (with a standard deviation equal to the effective lattice constant) has been applied to the individual stress components before the von Mises stress was calculated.}
    \label{fig:vonmises}
\end{figure}

Figure \ref{fig:vonmises} shows a comparison of the von Mises stress field between atomistic data and the continuum model for the case of an $\alpha'$ precipitate in an $\alpha$ matrix at 1000 K. In the left column of Figure \ref{fig:vonmises}, the continuum model and the atomistic data show comparable von Mises stress fields, but there is a noticeable discrepancy in the center of the precipitate. The von Mises stress according to the continuum model is zero, while it is approximately 0.2 GPa according to the atomistic data. Given that the eigenstrain is expected to be purely dilational, the deviatoric stress at the center of the precipitate should be equal to zero. We believe that due to the subsequent square and square-root operations of the stress fields in Eq. \ref{eq:20}, the von Mises stress is sensitive to noise, and that the non-zero magnitude in the center of the precipitate is a result of the magnitude of residual noise in the atomistic data. To test this, we apply a Gaussian filter (with a standard deviation equal to the effective lattice constant) to the individual stress components obtained from both the continuum model and the atomistic data before calculating the von Mises stress field. The second column of Figure \ref{fig:vonmises} shows the resulting fields after applying the Gaussian filter. As clearly shown in Figure \ref{fig:vonmises} (f), the largest effect of the filter on the atomistic data is to reduce the magnitude of the von Mises stress at the center of the precipitate. Even with 500 atomic snapshots, the residual noise is still enough to affect a reliable measurement of the average von Mises stress at the center of the precipitate. An accurate continuum model can circumvent issues of limited sampling sizes to determine an average stress field. Comparisons of von Mises stress fields for other situations are given on pages 7-10 of the Supplementary Material.

\subsection{Elastic Energies}

For the four different configurations investigated in this study ($\alpha'$ precipitate with a radius of approximately 5.8 nm in a $\alpha$ matrix at 600 K, $\alpha$ precipitate with a radius of approximately 5.7 nm in a $\alpha'$ matrix at 600 K, $\alpha'$ precipitate with a radius of approximately 5.3 nm in a $\alpha$ matrix at 1000 K, and $\alpha$ precipitate with a radius of approximately 5.6 nm in a $\alpha'$ matrix at 1000 K), we estimate the total elastic energy from the continuum model and the atomistic data. The elastic energy for the continuum model is given by Eq. \ref{eq:9}. As strain is difficult to determine from atomistic simulations with chemical heterogeneity, we estimate the elastic energy from atomistic data using the following equation:
\begin{equation} \label{eq:21}
    E^{el} = \frac{1}{2} \int d\vec{r} S_{ijkl}(\vec{r})\sigma_{ij}(\vec{r})\sigma_{kl}(\vec{r})
\end{equation}
where $S_{ijkl}(\vec{r})$ is the compliance tensor, which is related to the elastic constant tensor through \cite{sutton2020physics}:
\begin{equation} \label{eq:22}
    C_{ijkl}(\vec{r})S_{klmn}(\vec{r}) = \frac{1}{2}(\delta_{im}\delta_{jn}+\delta_{in}\delta_{jm}).
\end{equation}
The compliance tensor can be determined as a function of the local composition from the averaged atomistic composition fields and the compositionally dependent elastic constants from Eq. \ref{eq:10}.

The elastic energies for the different configurations are given in Table \ref{tab:elastic_energy}. The values between continuum modeling and atomistic data are in good agreement, with the largest discrepancy being a difference of about 15 \%.

\begin{table}[h!]
    \centering
	\begin{tabular}{l | l | l | l | l }
        \multirow{2}{*}{} 
         & $\alpha'$ in $\alpha$ (600 K) & $\alpha$ in $\alpha'$ (600 K) & $\alpha'$ in $\alpha$ (1000 K) & $\alpha$ in $\alpha'$ (1000 K) \\ \hline &&&& \\
		Continuum (eV) & 63.4 & 70.5 & 17.1 & 21.5 \\ &&&& \\ \hline &&&& \\
		Atomistic (eV) & 62.8 & 69.0 & 19.8 & 20.7 \\ &&&& \\

    \end{tabular}
	\caption{\label{tab:elastic_energy} Estimated total elastic energies of four precipitate/matrix configurations from atomistic data and from continuum modeling. The radii of the precipitates for the different configurations are, from left to right, 5.8 nm, 5.7 nm, 5.3 nm, and 5.6 nm.}
\end{table}

\subsection{Elastic Contributions to the Gibbs-Thompson Effect}

As an exercise to illustrate the importance of elastic effects on precipitation, we will investigate the importance of elasticity in estimating the interfacial free energy from the Gibbs-Thompson effect. If we assume that stress-composition coupling can be neglected, we can write the semi-grand potential ($\Omega (R)$) of a system with a spherical precipitate of radius $R$ as:
\begin{equation} \label{eq:23}
    \Omega(R) = \frac{4}{3}\pi R^3 \rho_0 \omega_0^p + (V-\frac{4}{3}\pi R^3) \rho_0 \omega_0^m + E^{el} + 4\pi R^2 \gamma
\end{equation}
where $\rho_0$ is the density, $\omega^p_0$ is the semi-grand potential per atom of the unstressed precipitate phase, $\omega^m_0$ is the semi-grand potential per atom of the unstressed matrix phase, $V$ is the total volume, and $\gamma$ is the interfacial free energy. By invoking the capillarity approximation and taking $d\Omega / dR = 0$, we obtain:
\begin{equation} \label{eq:24}
    0 = \rho_0(\omega^p_0 - \omega^m_0) + \frac{\partial E^{el} / \partial R}{4\pi R^2} + \frac{2\gamma}{R}.
\end{equation}
Following a derivation analogous to that found for Eq. 7.49 in Ref. \cite{Voorhees2004}, we use the following relations given that the exchange potential is constant everywhere at equilibrium:
\begin{equation} \label{eq:25}
    \omega_0^p = \mu^p_A, \omega_0^m = \mu^m_A
\end{equation}
\begin{equation} \label{eq:26}
    \mu^p_B - \mu^p_A = \mu^m_B - \mu^m_A
\end{equation}
and expand the chemical potentials for each phase to first order in composition about the equilibrium composition of each phase to obtain:
\begin{equation} \label{eq:27}
    c^p - c^p_e = \frac{(\partial E^{el} / \partial R)/(4\pi R^2) + 2\gamma/R}{\rho_0(c^p_e-c^m_e)f^{\prime \prime}_{p,e}}
\end{equation}
where $c^p$ is the composition of the precipitate with radius $R$, $c^p_e$/$c^m_e$ is the composition of the precipitate/matrix phase at coexistence, and $f^{\prime \prime}_{p,e}$ is the second derivative of the free energy with respect to composition of the precipitate phase at the coexistence composition. Eq. \ref{eq:27} describes how the composition of the precipitate can deviate from the coexistence composition based on contributions due to the interfacial free energy and elastic effects. 

We use Eq. \ref{eq:27} to make an estimate for the interfacial free energy of the precipitate. The left-hand side can be obtained by measuring the average composition of a precipitate away from the interface from atomistic simulations, and $f^{\prime \prime}_{p,e}$ can be estimated by the second derivative of Eq. \ref{eq:4}. The partial derivative of the elastic energy is estimated by:
\begin{equation} \label{eq:28}
    \frac{\partial E^{el}}{\partial R} \approx \frac{E^{el}(R+q)-E^{el}(R-q)}{2q}
\end{equation}
where we set $q$ to 0.1 nm. With information obtained from atomistic simulation and continuum elasticity modeling, we can determine the role of elastic effects on estimating the interfacial free energy.

Table \ref{tab:gt_test} shows the change in composition of the precipitate, second derivative of the free energy of the precipitate, and the estimated interfacial free energies when ignoring elastic effects and when considering elastic effects. When elastic effects are considered, the estimated interfacial free energy decreases by approximately 15 \% to 30 \%. Such a difference can significantly impact predictions of precipitation kinetics \cite{Tavenner2024}. For both 600 K and 1000 K, the difference between lattice constants of the two phases is less than 1 \%, and previous studies estimating the interfacial free energy of the Fe-Cr system have neglected elastic contributions due to the small lattice parameter difference between the two phases \cite{Sadigh2012_int,Tissot2023}. Even though an estimate of $\gamma$ from Eq. \ref{eq:27} can be considered crude due to the assumptions involved, the results in Table \ref{tab:gt_test} illustrate that elastic effects matter even for systems with relatively low misfit.

\begin{table}[h!]
    \centering
	\begin{tabular}{l | l | l | l | l }
        \multirow{2}{*}{} 
         & $c^p-c^p_e$ (\%) & $f^{\prime \prime}_{p,e}$ & $\gamma$ (eV/nm$^2$) & $\gamma$ (eV/nm$^2$)  \\ 
          & & (eV/atom) & w/o elastic effects & w/ elastic effects  \\ \hline &&&& \\
		$\alpha'$ in $\alpha$ (600 K) & 0.16 & 2.645 & 0.899 & 0.730  \\ &&&& \\ \hline &&&& \\
            $\alpha$ in $\alpha'$ (600 K) & -0.13 & 3.381 & 0.902 & 0.697  \\ &&&& \\ \hline &&&& \\
            $\alpha'$ in $\alpha$ (1000 K) & 1.01 & 0.202 & 0.304 & 0.246  \\ &&&& \\ \hline &&&& \\
		$\alpha$ in $\alpha'$ (1000 K) & -0.11 & 3.032 & 0.519 & 0.455  \\ &&&& \\

    \end{tabular}
	\caption{\label{tab:gt_test} Compositional shifts of precipitates, the second derivative of the free energy of the precipitate phase, and the estimated interfacial free energy using Eq. \ref{eq:27} with and without elastic effects.}
\end{table}

\section{Discussion} \label{discussion}

In this study, we have shown how finite-temperature properties obtained from atomistic modeling can inform a continuum elasticity model to approximate a finite-temperature atomistic stress field distribution. The advantage of using the continuum model is that the elastic energy of a system and its stress field can be easily estimated without relying on the generation and averaging of many atomistic snapshots. We have also shown that even in systems with small lattice misfit, elastic effects can still be significnat when considering the Gibbs-Thompson effect. While there is good agreement in general between the compared pressure and von Mises stress fields, there are some discrepancies present which may be exacerbated for systems with larger mismatch. These can be seen in the near-interface fields in Figures \ref{fig:press} and \ref{fig:vonmises}. Here we review the assumptions that have been made that are likely to be important in systems with higher lattice misfit.

Eq. \ref{eq:9} assumes linear elasticity, which is based on the small strain assumption. Higher-order elastic constants, or the consideration of strain-gradient elasticity may be necessary when this assumption is not valid. Additionally, the elastic constants (Eq. \ref{eq:10}) and the eigenstrains (Eq. \ref{eq:13}) are assumed to be linearly dependent on composition. Higher order dependencies could be relevant in other systems. There is also no direct consideration of the effect of interfacial tension, which could modify the stress field particularly for smaller precipitate sizes \cite{Cahn1982}.

We have also assumed that the precipitate has a spherical shape for the continuum model. Even if the interfacial free energy is isotropic, the equilibrium shape of the precipitate will deviate from that of a sphere at large sizes due to elastic effects \cite{Voorhees2004, Thompson1994,Boussinot2010}. In addition, if the interfacial free energy is anisotropic this would also change the precipitate shape. With a change in precipitate shape, the elastic fields would also change.

The elastic constants used in this study were obtained from an average of a few representative structures obtained from VC-SGC MC/MD. Ref. \cite{Mishin2015} showed that there is a difference in elastic constants at the same composition when obtained from MC in the canonical ensemble (a ``closed" system) compared to the SGC ensemble (an ``open" system). Depending on the magnitude of the difference, the choice of ensemble used when determining the elastic constants could affect the magnitude of the predicted stress-fields.

In addition to investigating the higher-order effects above, an additional future direction would be to compare the stress fields obtained from atomistic simulations and continuum modeling of a precipitate at or near a grain boundary. Previous phase field simulations have shown that the difference in orientation of two grains can dramatically vary the stress field of a precipitate in an elastically anisotropic material \cite{Heo2013}. As the previous study ignores the contribution of grain boundary elastic constants, comparison with atomistic results will unravel the importance of grain boundary elastic constants in determining a precipitate's stress field near a grain boundary.

\section{Conclusion} \label{conclusion}

We have obtained averaged finite-temperature stress fields of coherent precipitates in a Fe-Cr system using both atomistic simulations and continuum modeling. The parameters input into the continuum model include finite-temperature elastic constants and lattice constants. We find good agreement between the pressure fields obtained from both approaches, as well as with the von Mises stress fields after a Gaussian filter is applied. We also find that neglecting the elastic effect when estimating an interfacial free energy from the Gibbs-Thompson effect will result in a considerable overestimate of the interfacial free energy. As the Fe-Cr system has a relatively small lattice mismatch, this effect is likely more pronounced in other systems that show coherent precipitation. Additionally, many of the assumptions used in this work may require further investigation when considering systems with higher mismatch.

\section*{Acknowledgements}

The authors acknowledge insightful discussions with Yuri Mishin and Lucas Hale. A.A. acknowledges support from a NRC Postdoctoral Fellowship while at the National Institute of Standards and Technology.






\end{document}